\documentclass[8.5pt,twoside,twocolumn]{article}
\usepackage[super,sort&compress,comma]{natbib} 
\usepackage{mhchem}
\usepackage{amsmath}
\usepackage{amssymb} 
\usepackage{amsthm}
\usepackage{times,mathptmx}
\usepackage{sectsty}
\usepackage{balance} 
\usepackage[usenames,dvipsnames]{color}
\usepackage{bm}
\usepackage{graphicx} 
\usepackage{lastpage}
\usepackage[format=plain,justification=raggedright,singlelinecheck=false,font=small,labelfont=bf,labelsep=space]{caption} 
\usepackage{fancyhdr}
\begin{document}

\thispagestyle{plain}
\fancypagestyle{plain}{
\fancyhead[L]{\includegraphics[height=8pt]{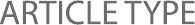}}
\fancyhead[C]{\hspace{-1cm}\includegraphics[height=20pt]{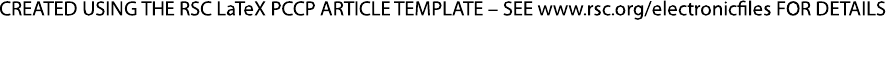}}
\fancyhead[R]{\includegraphics[height=10pt]{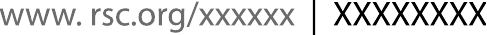}\vspace{-0.2cm}}
\renewcommand{\headrulewidth}{1pt}}
\renewcommand{\thefootnote}{\fnsymbol{footnote}}
\renewcommand\footnoterule{\vspace*{1pt}%
\hrule width 3.4in height 0.4pt \vspace*{5pt}} 
\setcounter{secnumdepth}{5}

\makeatletter 
\def\subsubsection{\@startsection{subsubsection}{3}{10pt}{-1.25ex plus -1ex minus -.1ex}{0ex plus 0ex}{\normalsize\bf}} 
\def\paragraph{\@startsection{paragraph}{4}{10pt}{-1.25ex plus -1ex minus -.1ex}{0ex plus 0ex}{\normalsize\textit}} 
\renewcommand\@biblabel[1]{#1}            
\renewcommand\@makefntext[1]%
{\noindent\makebox[0pt][r]{\@thefnmark\,}#1}
\makeatother 
\renewcommand{\figurename}{\small{Fig.}~}
\sectionfont{\large}
\subsectionfont{\normalsize} 

\fancyfoot{}
\fancyfoot[LO,RE]{\vspace{-7pt}\includegraphics[height=9pt]{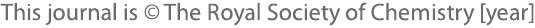}}
\fancyfoot[CO]{\vspace{-7.2pt}\hspace{12.2cm}\includegraphics{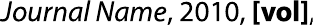}}
\fancyfoot[CE]{\vspace{-7.5pt}\hspace{-13.5cm}\includegraphics{RF}}
\fancyfoot[RO]{\footnotesize{\sffamily{1--\pageref{LastPage} ~\textbar  \hspace{2pt}\thepage}}}
\fancyfoot[LE]{\footnotesize{\sffamily{\thepage~\textbar\hspace{3.45cm} 1--\pageref{LastPage}}}}
\fancyhead{}
\renewcommand{\headrulewidth}{1pt} 
\renewcommand{\footrulewidth}{1pt}
\setlength{\arrayrulewidth}{1pt}
\setlength{\columnsep}{6.5mm}
\setlength\bibsep{1pt}

\twocolumn[
  \begin{@twocolumnfalse}
\noindent\LARGE{\textbf{Rings in Random Environments: Sensing Disorder Through Topology}}
\vspace{0.6cm}

\noindent\large{\textbf{Davide Michieletto\textit{$^{a}$},
Marco Baiesi{$^{b,c}$},
Enzo Orlandini\textit{$^{b,c}$}, 
Matthew S. Turner{$^{\ast}$}{$^{a}$}}}\vspace{0.5cm}

\noindent\textit{\small{\textbf{Received Xth XXXXXXXXXX 20XX, Accepted Xth XXXXXXXXX 20XX\newline
First published on the web Xth XXXXXXXXXX 200X}}}

\noindent \textbf{\small{DOI: 10.1039/b000000x}}
\vspace{0.6cm}

\noindent \normalsize{In this paper we study the role of topology in DNA gel electrophoresis experiments via molecular dynamics simulations. The gel is modelled as a 3D array of obstacles from which half edges are removed at random with probability $p$, thereby generating a disordered environment. Changes in the microscopic structure of the gel are captured by measuring the electrophoretic mobility of ring polymers moving through the medium, while their linear counterparts provide a control system as we show they are insensitive to these changes. We show that ring polymers provide a novel, non-invasive way of exploiting topology to sense microscopic disorder. Finally, we compare the results from the simulations with an analytical model for the non-equilibrium differential mobility, and find a striking agreement between simulation and theory.}
\vspace{0.5cm}
 \end{@twocolumnfalse}
  ]

\section{Introduction}

\footnotetext{\textit{$^{a}$ Department of Physics and Centre for Complexity Science, University of Warwick, Coventry CV4 7AL, United Kingdom.}}
\footnotetext{\textit{$^{b}$ Dipartimento di Fisica e Astronomia, Universit\`a di Padova, Via Marzolo 8, 35131 Padova, Italy.}}
\footnotetext{\textit{$^{c}$ INFN, Sezione di Padova, Via Marzolo 8, 35131 Padova, Italy.}}
\footnotetext{${\ast}$ corresponding author.}

The transport properties of circular polymers moving in crowded environments has been studied for several decades~\cite{DeGennes1971,Doi1988,Klein1986,Rubinstein1986,Alon1997,Cole2002,Rosa2011,Michieletto2014,Michieletto2014c},  but a complete understanding remains elusive. As well as providing an interesting problem in theoretical physics in which topology plays a role, the dynamics of ring polymers in
complex environments is a problem that has important practical applications. 

Gel electrophoresis, a ubiquitous technique~\cite{Calladine1997,Viovy2000,Cole2006,Dorfman2010}, relies on the fact that polymers with different molecular weight, length or topology migrate at different speeds when forced to move through an intricate medium, like an agarose gel, by the action of an electric field~\cite{Mickel1977,Levene1987,Stasiak:1996:Nature:8906784,Trigueros2001}.  This allows the separation of polymers with different physical properties and is systematically used for DNA identification and purification~\cite{Calladine1997}. However, relatively little is known about how ring polymers move through a real gel~\cite{Calladine1991,Alon1997,Viovy2000,Cole2002,Cole2003a}.

Previous theoretical models have often treated gels as a perfect mesh of  obstacles~\cite{Rubinstein1986,Cates1986,Calladine1991,Michieletto2014,Michieletto2014c}. 
On the other hand, it is well known that physical gels have irregularities, such as
dangling ends~\cite{Whytock1991,Cole2003a,Rahong2014}. These are more common in agarose gels formed at low agarose concentrations because many of the agarose bundles fail to cross-link with other fibers, thereby generating partially cross-linked open strands\cite{Whytock1991}. More recently, dangling ends have also been directly observed in artificial gels made of solid nano-wires using transmission electron microscopy~\cite{Rahong2014}.

\begin{figure*}[t]
\includegraphics[scale=0.1]{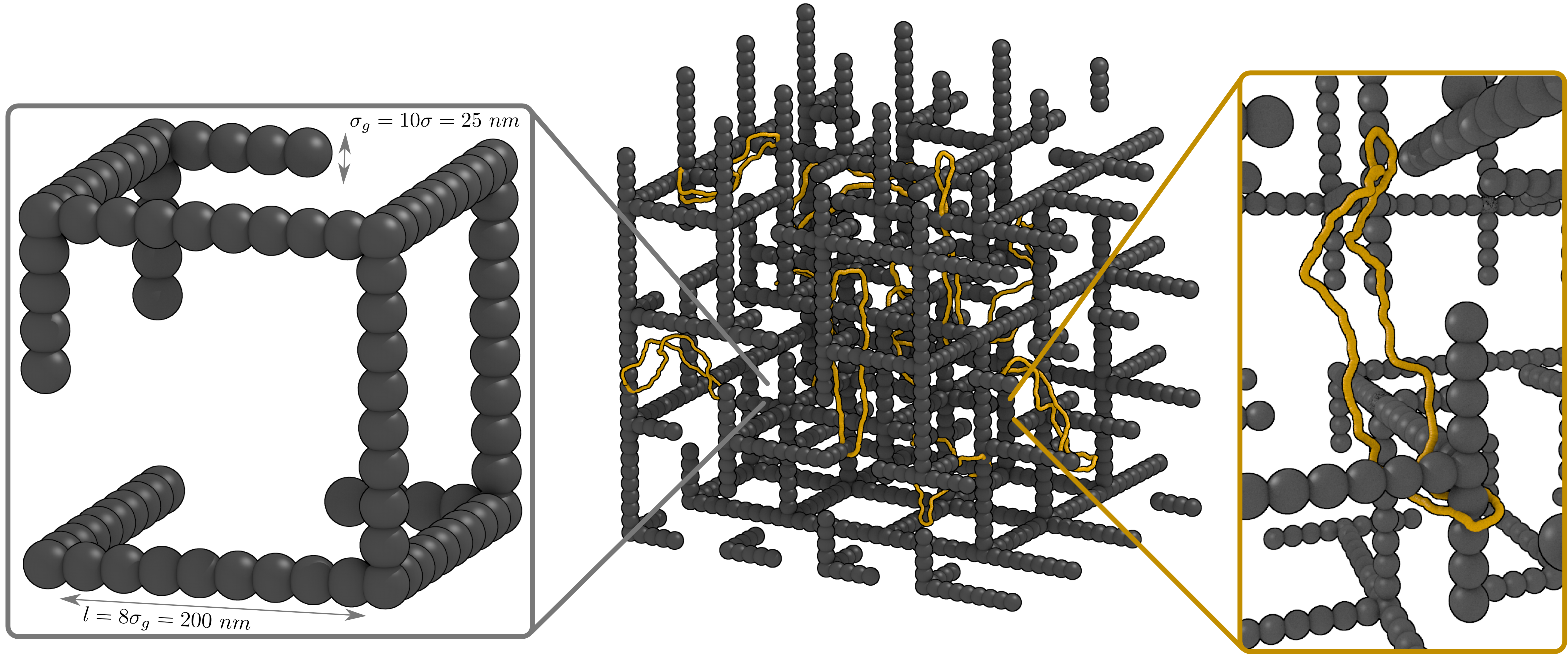}
\caption{Assembling a random environment: With probability $p$ half edge is knocked out in a random direction. By assembling multiple cells we construct a random environment which strongly resembles the disordered structure of a gel, where dangling ends populate the medium. To the far right we show the case where a ring is impaled by a dangling end. The external field in this case is directed upward. Periodic boundary conditions are applied to the simulation box. The gel structure is here thinned for clarity.}
\label{fig:Panel1}
\end{figure*}

The presence of these dangling ends plays a very weak role when linear polymers are undergoing gel electrophoresis. Conversely, gel electrophoresis experiments involving polymers with looped structures are expected to depend rather strongly on the topological interactions between the polymers and the gel structure (see Fig.~\ref{fig:Panel1}). When a dangling end threads through the ring polymer, the latter becomes ``impaled'' and its free motion is re-established only when the threading is removed. This makes the presence of dangling ends in the gel a crucial element for any realistic model aiming to faithfully describe gel electrophoresis dynamics of  polymers with an architecture that includes one or more closed loops.

Here we study the dynamical properties of charged rings subject to an external electric field that move within a mesh of obstacles that is regular except for the presence of randomly positioned dangling ends. These defects strongly interact with the topology of the rings and these interactions may give rise to striking, counter-intuitive behaviour.

In particular we show that, in the regime of strong electric field and sufficiently high concentration of dangling ends, rings migrate slower as the external bias (field) is further increased. This gives rise to negative differential mobility~\cite{Zia2002,Baerts2013,Ghosh2014}. The phrase ``getting more from pushing less'' has often been used to describe this behaviour, where the flux generated by external fields is smaller for stronger fields than for weaker ones. The system we study therefore represents an interesting new example of negative differential mobility. Later in tis article we exploit some recent theoretical results to relate the source of this unusual behaviour to the correlation between forces and displacements.

The topological interactions between ring polymers and the gel architecture can also provide information on the microscopic structure of the gel. Our results suggest that a gel electrophoresis experiment can establish the level of disorder in the medium by comparing the results obtained by running linear and ring polymers. This represents a novel way to exploit topology preservation to ``sense'' the disorder in the microscopic structure of a material in a non-invasive way.

The paper is structured as follows: In Sec.~\ref{s:model} we introduce the model and describe the computational details. In Sec.~\ref{s:results} we present our findings and compare them with analytic predictions. In Sec.~\ref{s:conclusions} we give our conclusions and discuss further the application of topology as a method for sensing the microscopic disorder in materials.

\section{Model and Methods}
\label{s:model}
The random gel through which the rings move is modelled as a regular cubic lattice in which a fraction $p$ of dangling ends is produced by halving randomly the edges of the cubic lattice with probability $p$. A typical configuration of this random mesh is reported in Fig.~\ref{fig:Panel1}. By properly tuning the value of $p$, random meshes with different average number of dangling ends can be obtained. The lattice spacing of the gel is $200$ $nm$ and has been  chosen to reproduce the pore size of an agarose gel at 5\%~\cite{Pernodet1997}. While this value is rather high for the gel used in standard experiments of gel electrophoresis, it is comparable to the typical gel lattice spacing involved in high resolution experiments where, for instance, polyacrylamide gel 
at 5\%T and 2.7\%C~\cite{Stellwagen2009} is considered. 
This value is also well within the range of pore-sizes accessible to arrays of solid nano-wires~\cite{Rahong2014}, which represents a candidate for ultra-fast gel electrophoresis. The mesh of the gel is here constructed as a cubic lattice made of static beads of diameter
$\sigma_g=10 \sigma$, where $\sigma$ is the size of the beads forming the polymer chains (see below). By taking $\sigma = 2.5$ $nm$, which is the typical diameter size of hydrated dsDNA, the thickness of the mesh is $25$ $nm$ which is compatible with the 
typical diameter of either agarose bundles ($30$ $nm$~\cite{Pernodet1997}) and nano-wires ($20$ $nm$~\cite{Rahong2014}). Note that the gel is assumed not to significantly deform due to both, thermal fluctuations or collision events with the beads of the moving polymers. This is something of an oversimplification for agarose or polyacrylamide gels, which have a relatively small Young's modulus, but it is quite reasonable for gels made of solid nano-wires, whose Young's modulus is $10^5$ times bigger than the agarose one~\cite{Rahong2014}. Our gel structure can therefore be interpreted as a close representation of the structure of solid nano-wires mentioned above, or a (zero order) approximation of an agarose or polyacrylamide gel. This assumption of rigidity has significant computational advantages in terms of algorithmic speed. 

Rings moving through the gel are modelled by a set of $N$ circular worm-like chains each of $M$ beads of 
diameter $\sigma=2.5$ $nm$.  Here we consider either systems of $N=10$ rings with $M=512$ beads 
each or systems of $N=20$ rings and $M=256$ beads. 
These two cases correspond to circular DNA of about $3.7$ $kbp$ (contour length $L_c \simeq 1.3$ $\mu m$) 
and $1.9$ $kbp$ ($L_c \simeq 0.65$ $\mu m$) respectively. The whole system is contained within a box of 
linear dimension $L=320\sigma$ and has periodic boundary conditions in all three directions.
In both cases, the system volume fraction is $\phi \simeq NM\sigma^3/(L^3 - V_{gel}(p)) \lesssim 1.6 \cdot 10^{-4}$, much smaller than the value at which the chains start to overlap ($\phi^{*} =33.3 \cdot 10^{-4}$). This means that the systems are in the dilute limit. Interactions between rings are therefore neglected in the analytic calculations, although they do occur (rarely) in the simulations.
In the following we shall indicate with $\bm{r}_i$ the position of the center of the $i$-th bead and 
with $\bm{d_{i,j}} \equiv \bm{r}_i - \bm{r}_{j}$ the virtual bond vector of length $d_{i,j}$ 
connecting beads $i$ and $j$. The connectivity of the chain is treated within the finitely extensible non-linear elastic model~\cite{Kremer1990} with potential energy defined as, 
\begin{equation}
U_{FENE}(i,i+1) = -\dfrac{k}{2} R_0^2 \ln \left[ 1 - \left( \dfrac{d_{i,i+1}}{R_0}\right)^2\right]  \notag
\end{equation}
for  $d_{i,i+1} < R_0$ and $U_{FENE}(i,i+1) = \infty$, otherwise; here we chose $R_0 = 1.6$ $\sigma$ and $k=30$ $\epsilon/\sigma^2$ and the thermal energy $k_BT$ is set to $\epsilon$. The bending rigidity of DNA is captured with a standard Kratky-Porod potential,
\begin{equation}
U_b(i,i+1,i+2) = \dfrac{k_BT \l_K}{2\sigma}\left[ 1 - \dfrac{\bm{d}_{i,i+1} \cdot \bm{d}_{i+1,i+2}}{d_{i,i+1}d_{i+1,i+2}} \right],\notag
\end{equation}
where $l_K = 40 \sigma \simeq 100 nm $ is the known Kuhn length of unconstrained DNA. The steric interaction between beads belonging to the polymers or the gel is taken into account by a truncated and shifted Lennard-Jones potential  
\begin{equation}
U_{LJ}(i,j) = 4 \epsilon \left[ \left(\dfrac{\sigma_c}{d_{i,j}}\right)^{12} - \left(\dfrac{\sigma_c}{d_{i,j}}\right)^6 + 1/4\right] \theta(2^{1/6}\sigma_c - d_{i,j}) \notag.
\end{equation} 
where $\theta(x)$ is the usual step function and $\sigma_c$ is the minimum distance between bead centres, \emph{i.e.} $\sigma_c = \sigma$ between polymer beads and $\sigma_c = (\sigma_g + \sigma)/2 = 5.5$ $\sigma$ between a polymer bead and a gel bead. The beads forming the mesh are placed $\sigma_g$ apart and steric interactions between themselves are excluded from the computation. 

Denoting by $U$ the total potential energy, the dynamic of the beads forming the rings is described within a Langevin scheme:
\begin{equation}
m \ddot{\bm{r}}_i = - \xi \dot{\bm{r}}_i - {\nabla U} + \bm{\eta}
\label{langevin}
\end{equation}
where $\xi$ is the friction coefficient and $\bm{\eta}$ is the
stochastic delta-correlated noise. The variance of each Cartesian component of the noise, $\sigma_{\eta}^2$ satisfies the usual 
fluctuation dissipation relationship $\sigma_{\eta}^2 = 2 \xi k_B T$.

As customary~\cite{Kremer1990}, we set $m/\xi = \tau_{LJ}$, with $\tau_{LJ} = \sigma \sqrt{m/\epsilon}= \sigma \sqrt{m/k_B T}$ being the characteristic simulation time. From the Stokes friction coefficient of spherical beads of diameters $\sigma$ we have: $\xi = 3 \pi \eta_{sol} \sigma$ where $\eta_{sol}$ is the solution viscosity. By using the nominal water viscosity, $\eta_{sol}=1$ $cP$ and setting $T=300$ K and $\sigma=2.5$ $nm$, one has $\tau_{LJ} = {3 \pi \eta_{sol} \sigma^3/\epsilon} = 37$ $ns$. Since we keep the gel structure static for all time no equation of motion is necessary for the beads belonging to the mesh.

The numerical integration of Eq.~\eqref{langevin} is performed by 
using a standard Verlet algorithm with time step $\Delta t = 0.01 \tau_{LJ}\sim 0.4$ $ns$. Assuming that the electric charge is uniformly distributed along the rings, the total force $\bm{F}$ acting on each ring is $\bm{F} = q_{r} \bm{E}$, where $q_{r} = M q$ and $q$ is the representative charge of a single bead. 
The force acting on each bead can be expressed in units of $\epsilon/\sigma \simeq 1.6 $ $pN$, the total force acting on the rings being $F = M \cdot 1.6 pN$. Since each bead corresponds to $\sigma = 2.5$ $nm \simeq 7$ $bp$, and each base-pair contains two phosphate groups which account for a negative charge each, one can approximate the charge in each bead as $14 q_e$, where $q_e$ is the electron charge. In this case, the force applied to the beads can be thought of as a result of the action of an external electric field pointed towards $-\hat{z}$ and re-scaled by the charge of a bead. This means that the field strength applied to each bead can be expressed in units of $V/cm$  as $\tilde{E} = 1.6 pN/22.43\cdot 10^{-19} C \simeq 700$ $V/cm$. In this work, the fields used range from $10^{-3}$ to $10^{-1} \tilde{E}$, \emph{i.e.} between $0.7$ and $70$ $V/cm$, which are compatible to the values used in standard DNA gel electrophoresis~\cite{Mickel1977,Levene1987,Viovy2000}.

The starting point of the various simulations is a configuration in which flat rings are placed outside the box. We then slowly pull the rings inside the gel structure, avoiding any impalement. Finally, we locate the boundaries of the simulation box in order to be commensurate with the gel structure and impose periodicity in all the three directions. We first pre equilibrate the system 
for $5 \cdot 10^6$ $\tau_{LJ}$ time steps before the electric field is switched on. The mean and standard error of all quantities of interest is calculated by averaging over all the rings in the system.

\section{Results}
\label{s:results}
\begin{figure}[t]
\includegraphics[scale=0.7]{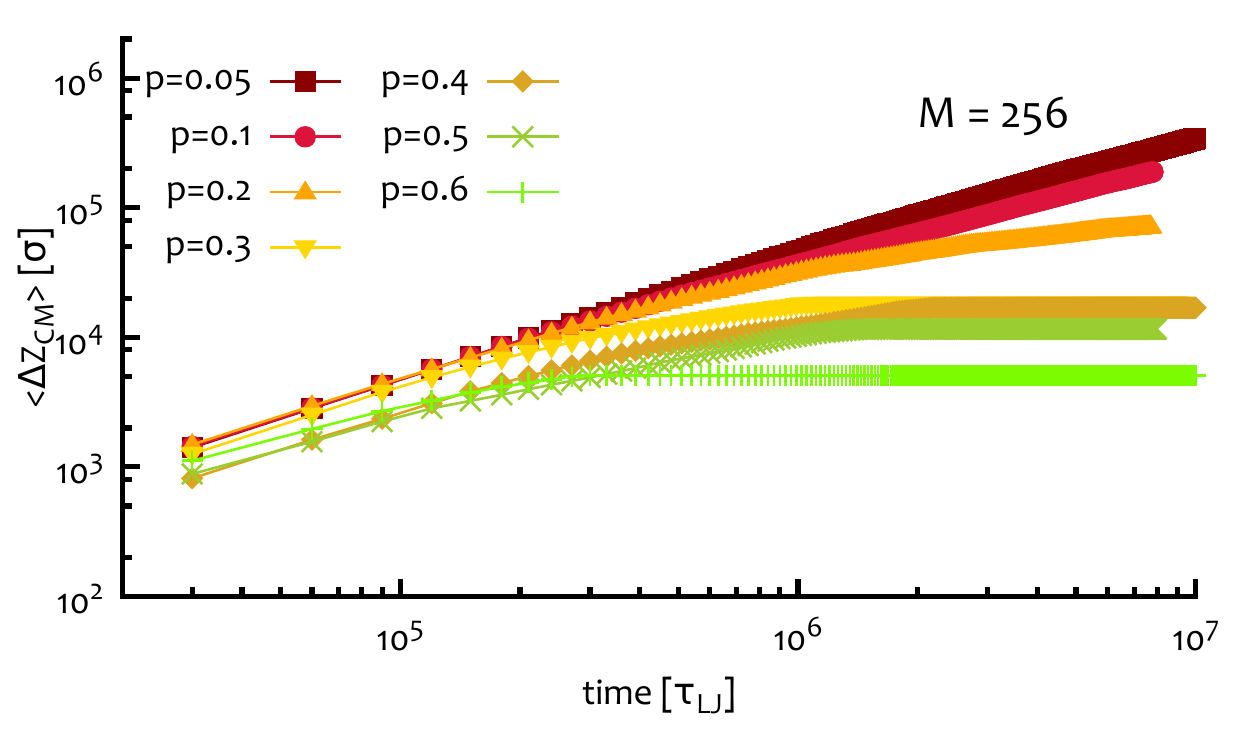}\\
\includegraphics[scale=0.7]{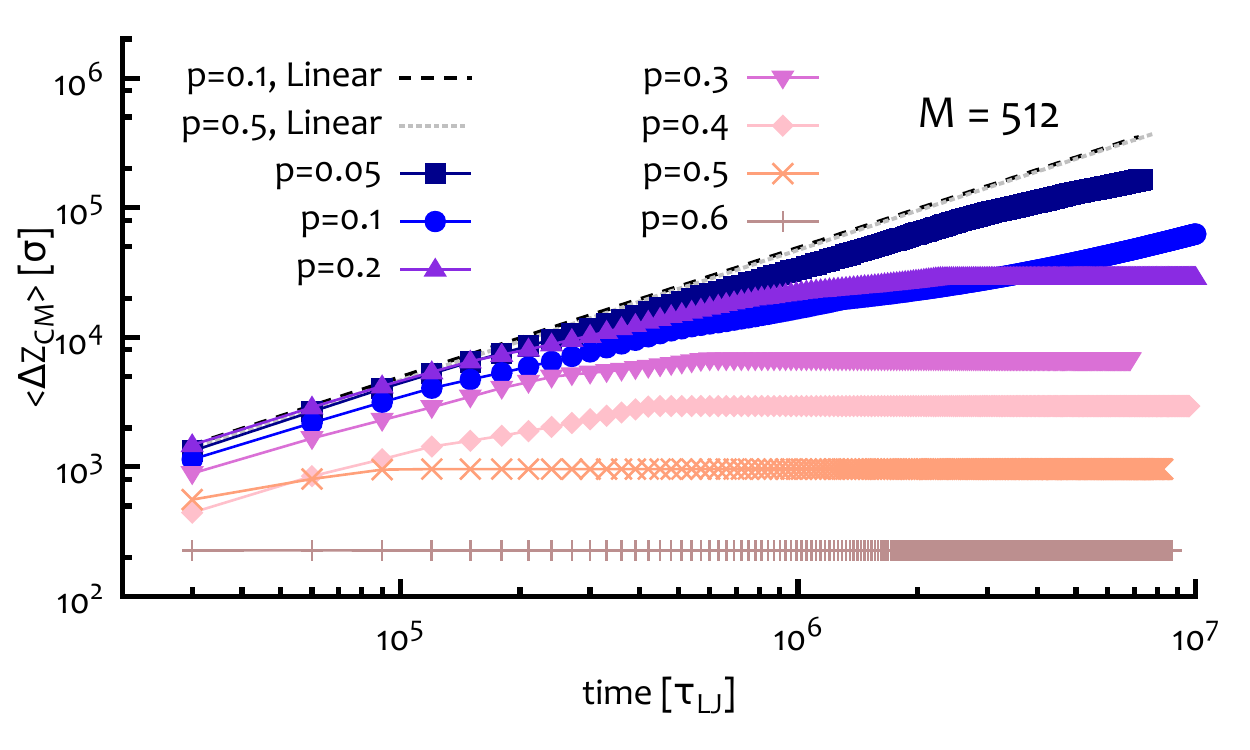}\\
\caption{Time dependence of the average centre of mass displacement 
along the $z$-direction, $\langle \Delta Z_{CM} \rangle$, of rings inside the random mesh. The external field strength is $E=0.05$ $\epsilon/q\sigma$ and the two sets of rings considered have $M=256$ (top) and $M=512$ (bottom) beads. Different curves refer to different values of $p$. Results for linear polymers with $M=512$ and two different values of $p$ are reported as dashed black and dotted grey lines in the bottom panel. In system units $10^6$ $\sigma \simeq 2.5$ $mm$ and $10^7$ $\tau_{LJ} \simeq 0.37$ $s$.}
\label{fig:ZcmDrift}
\end{figure}

To address the effect that the presence of dangling ends has 
on the dynamics of charged rings in the presence of an external electric field $\bm{E}$, we first monitor the average centre of mass displacement along the direction $\hat{z}$ of the field, $\langle \Delta Z_{CM} \rangle$ for different values of the parameter $p$ (see Fig.~\ref{fig:ZcmDrift}). It is apparent that the rings significantly slow down as the fraction  $p$ of incomplete edges (dangling ends) increases. To confirm that this is due to the ring topology we repeated the simulations for linear chains with the same contour lengths (see black dashed and grey dotted lines in Fig.~\ref{fig:ZcmDrift}): as expected the dangling ends do not interfere with the motion of the linear chains that is broadly insensitive to changes in the microscopic structure of the gel (i.e. changes in $p$).

From this data, we can estimate the rings mobility $\mu(M,p)$ as a function of rings size and probability of creating a dangling end $p$, by the relation  
\begin{equation}
\mu(M,p) = \dfrac{\langle v \rangle }{\left| \bm{F} \right|}
\label{eq:mobility}
\end{equation}
where the average velocity $\langle v \rangle$ is computed
from the long time limit of the centre of mass displacement, namely
\begin{equation}
\langle v \rangle = \lim_{t\rightarrow \infty} \dfrac{\langle \Delta Z_{CM}(t) \rangle}{t}.
\label{eq:vel}
\end{equation}

\begin{figure}[t]
\includegraphics[scale=0.7]{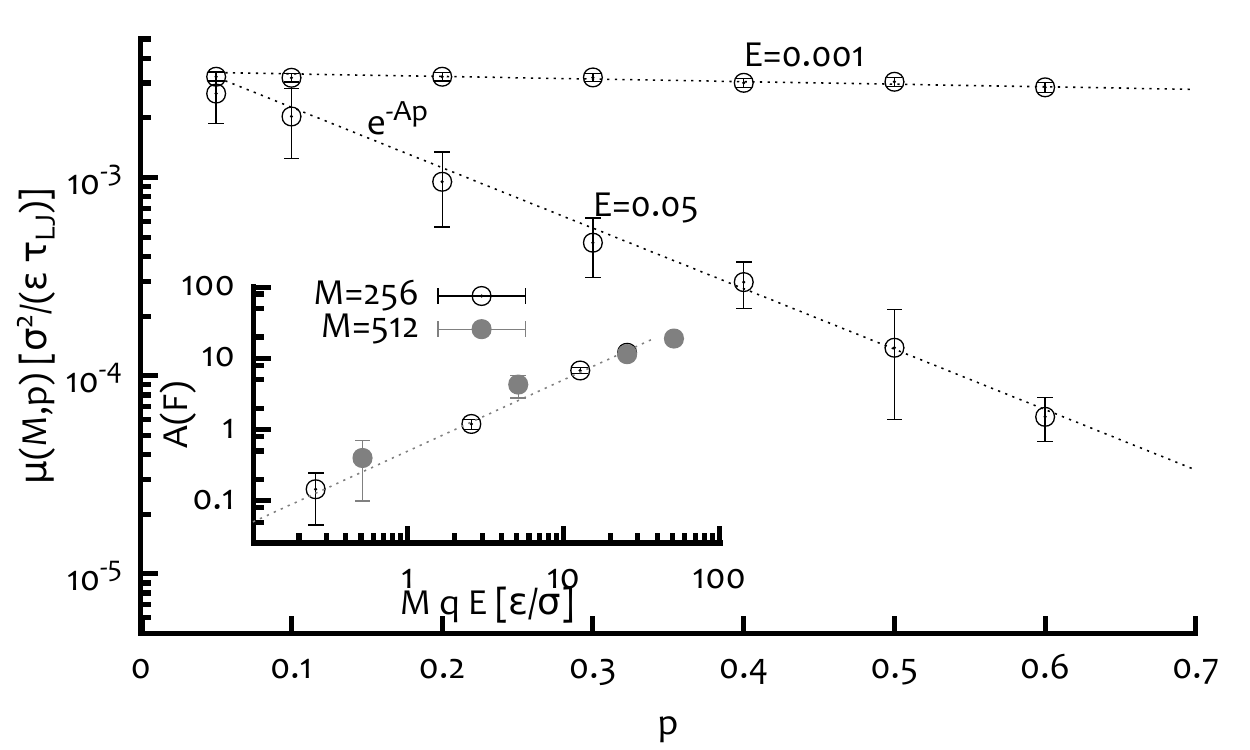}\\
\caption{Rings mobility $\mu$, as defined in Eq.~\eqref{eq:mobility}, as a function of the average fraction of dangling ends $p$. The rings considered have $M= 256$ and the two sets of symbols refer to two values of the field strength $E$. The data are well fitted by exponentials of the form $\mu \sim e^{-Ap}$ (dotted lines). In the inset, we show that the pre-factor $A$ seems to scale extensively with $F=MqE$ (see text).}
\label{fig:Mobility1}
\end{figure}


In the main panel of Fig.~\ref{fig:Mobility1} we report $\mu(M,p)$ for rings of 
$M=256$ beads driven by two  different external fields strengths $\left| \bm{E} 
\right| = E$, as a function of the average fraction of dangling ends $p$. As 
expected, for fixed $E$ and $M$, the mobility decreases as $p$ increases since 
rings will be more often impaled by the dangling ends. This behaviour is 
apparently well captured by an exponential law of the form
$\mu \sim e^{-Ap}$, where $A$ seems to scale extensively with $MqE$ (inset of Fig.~\ref{fig:Mobility1}). One can interpret this result within the assumption that the rings move with mobility $\mu_0$ only when not impaled and otherwise are essentially immobile. Hence
\begin{equation}
\mu = (1-s) \mu_0
\label{eq:mobp}
\end{equation}
with $s$ the fraction of time in which a ring is stuck. This may be 
approximated by $s=\alpha p v /(\alpha p v + e^{-\Delta G/k_BT})$, where the 
rate of hitting a dangling end in the mobile state is proportional to both the 
velocity $v$ and the density of ends $p$ and the disentanglement rate is 
Arrhenius-like and proportional to $e^{-\Delta G/k_BT}$ with $\Delta G$ the 
relevant free energy barrier for disengagement from a dangling end. The 
fluctuation of energy $\Delta G$ required to disentangle from an impaled 
situation may be expected to have Arrhenius form $\Delta G =M q E l/2$ where the 
ring must move the length of the penetrating segment $l/2$ against a force 
$F=MqE$. For strong fields this captures the exponential variation of $A\sim 
MqE$ but not the exponential variation with $p$. Although we do not have a simple explanation for the latter, this might be related to the higher order structure of ramified ring polymers which is completely neglected in our analysis.

\begin{figure}[t]
\includegraphics[scale=0.7]{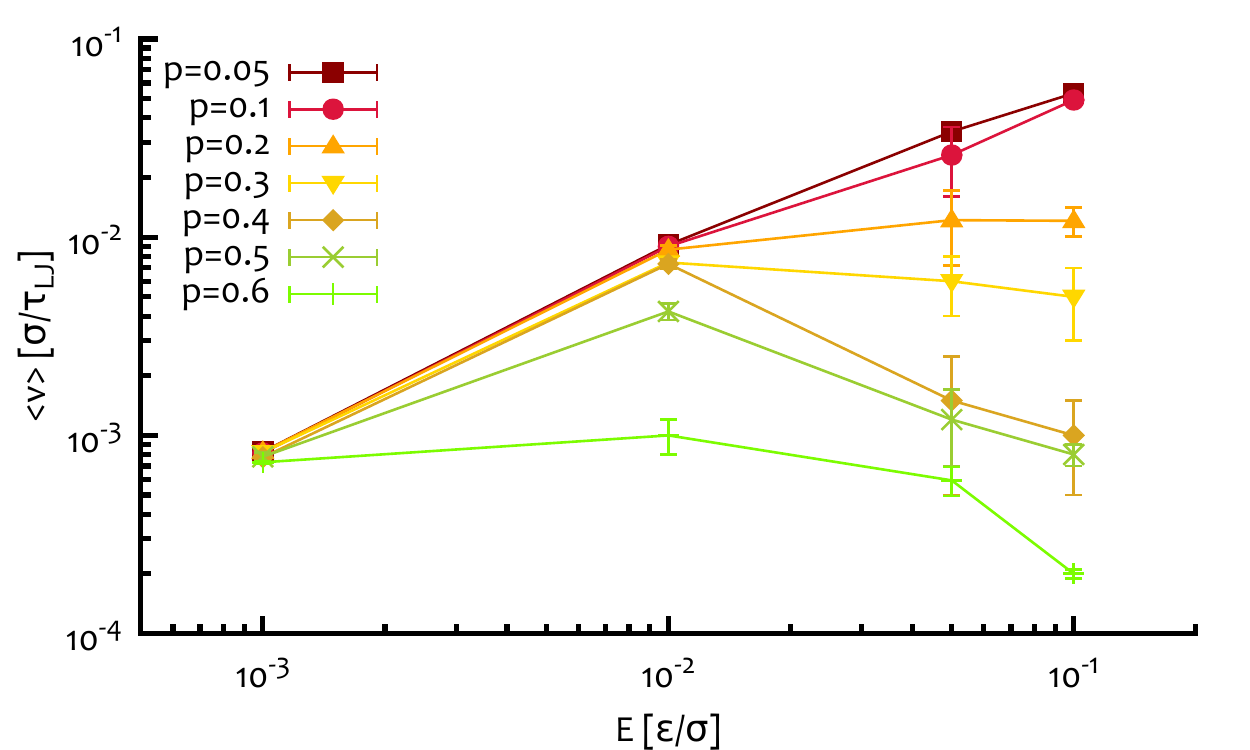}\\
\includegraphics[scale=0.7]{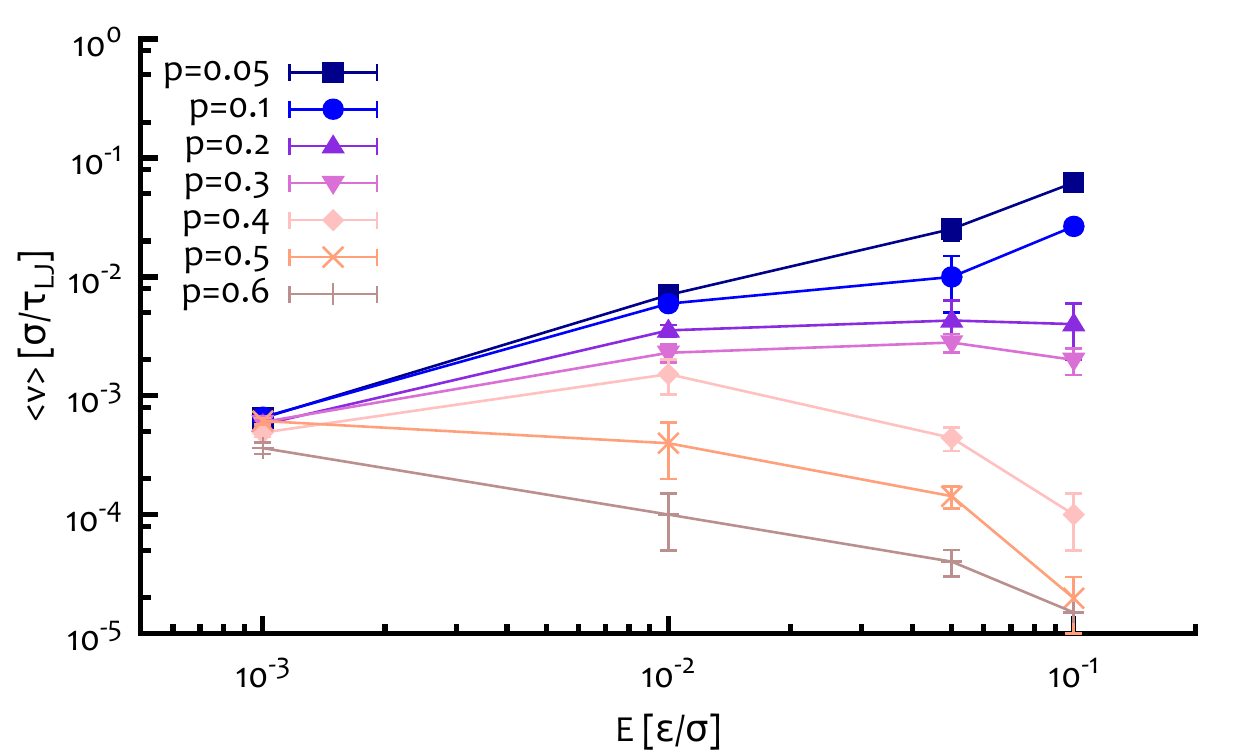}\\
\caption{The average speed $\langle v \rangle$ of ring polymers as a function of the force on the single beads $qE$ for rings with $M=256$ (top) and $M=512$ (bottom) beads is shown. Notice the clear negative differential mobility $\partial \langle v \rangle / \partial E$ for fields $E>0.01$ and $p > 0.3$. In our system units $0.1$ $\sigma/\tau_{LJ} \simeq 6.7$ $mm/s$ and $0.01$ $\epsilon/q\sigma \simeq 7$ $V/cm$.}
\label{fig:VelvsField}
\end{figure}

Fig.~\ref{fig:VelvsField} reveals the striking non-equilibrium property that the average centre of mass speed $\langle v \rangle$ decreases as the strength of the field $E$ increases for sufficiently high density of dangling ends $p$. 
This qualitative behaviour can be related to the definition of the mobility in eq.~\eqref{eq:mobility}  combined with the empirical result of Fig.~\ref{fig:Mobility1} that the mobility decreases with force. The average velocity might then be expected to scale as $\langle v \rangle \sim F\>e^{-c p F}$, with $c$ a constant, which has a maximum at an intermediate value of $F$. 

Following previous studies on the the linear response in non-equilibrium systems~\cite{Baiesi2009c,Baiesi2011,Baerts2013} we can estimate the differential mobility of the rings
\begin{equation}
\mu_D^N = \dfrac{\partial}{\partial F} { \langle v (F) \rangle}
\label{eq:DiffMobN}
\end{equation} 
as the long time limit of the response of the mean velocity to a change in the field strength $E$. In Appendix  we show that this has the following form, 
\begin{equation}
\mu_D^A = \lim_{t\rightarrow \infty} \dfrac{\mathcal{D}(t) - \mathcal{C}(t)}{k_BT},
\label{eq:DiffMobA}
\end{equation}
where the first term
\begin{equation}
\mathcal{D}(t) = \dfrac{1}{2t} \left[ \langle \Delta z^2(t) \rangle - \langle \Delta z(t)\rangle^2 \right]
\label{eq:noneqdiff}
\end{equation}
is the nonequilibrium generalisation of the 1D diffusion constant, and the second term is
\begin{equation}
\mathcal{C}(t) =  \dfrac{1}{2t} \left[ \left\langle \int_0^t ds \nu \Psi(s) \Delta z(t) \right\rangle  - \left\langle \int_0^t ds \nu \Psi(s) \right\rangle \left\langle \Delta z(t) \right\rangle \right],
\label{eq:noneqc}
\end{equation}
\begin{figure}[t]
\includegraphics[scale=0.7]{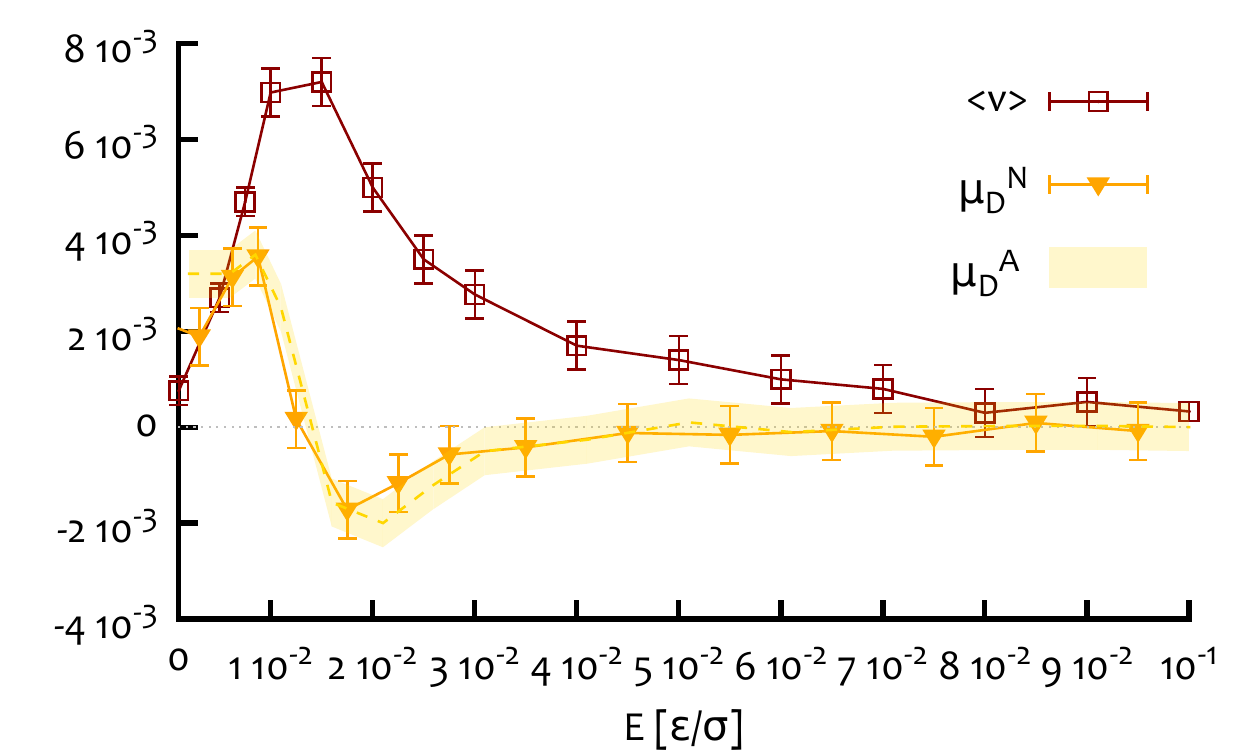}\\
\caption{Average speed of the centre of mass of rings, $\langle v \rangle$ in units of $\sigma/\tau_{LJ}$, for a system of $N=20$ rings with $M=256$ beads each and with $p=0.4$ (dark-yellow diamonds in Fig.~\ref{fig:VelvsField}). The differential mobilities computed as in eq.~\eqref{eq:DiffMobN} $\mu^N_D$ and as in eq.~\eqref{eq:DiffMobA} $\mu^A_D$, in units of $\sigma^2/\epsilon\tau_{LJ}$, as a function of the force acting on single beads $qE$ are shown to be in very good agreement. In system units the minimum of the differential mobility is reached at field strength around $0.02$ $\epsilon/q\sigma \simeq 14$ $V/cm$, which is compatible with experiments~\cite{Mickel1977}.}
\label{fig:DiffMob}
\end{figure}
where $\nu = \xi^{-1}$ is the inverse of the friction acting on each polymer bead, $\Delta z(t)$ is the displacement of a bead at time $t$ along $\hat{z}$ and $\Psi(s) = -\partial U(s)/\partial z + F$ is the sum of all the forces projected on $\hat z$ acting on a bead at time $s$~\cite{Baiesi2009c,Baiesi2011}. This covariance between time-averaged forces and displacements appears because the rings are pushed significantly far from equilibrium via an external force $F = MqE$. Clearly at equilibrium detailed balance holds, the displacements are on average  uncorrelated with the force, $\mathcal{C}(t)$ would be zero and the fluctuation-dissipation theorem would be valid according to the usual form of the Einstein relation $\mu = \mathcal{D} / k_BT$. 

It is interesting to notice that the presence of the non-equilibrium term $\mathcal{C}$ may introduce novel features into the system. In particular, from Eq.~\eqref{eq:DiffMobA}, it is apparent that, if $\mathcal{C}>\mathcal{D}$, the differential mobility becomes negative. This is an exclusive aspect of non-equilibrium systems. In order to compute $\mu_D^A$, we first reach a steady state with constant average velocity at a fixed $E$ and then we slightly increase the field strength. We then monitor the position and the sum of all the forces acting on the beads at each time-step until a new steady state is reached. The average is then performed over the beads belonging to a ring and over all the rings of the system. 

The differential mobility can also be computed numerically by taking the differential of the velocity with respect to the applied force, \emph{i.e.} through a discrete derivative $\mu_D^N = \Delta \langle v \rangle /\Delta F $ of Eq.~\eqref{eq:DiffMobN}. To evaluate this derivative, we make use of the average speed $\langle v \rangle$ computed in Fig.~\ref{fig:VelvsField} and Eq.~\eqref{eq:DiffMobN}. In Fig.~\ref{fig:DiffMob} we show the results obtained via these two different approaches for the case of a system with rings with length $M=256$ beads and gel with $p=0.4$. As one can notice, the agreement is very good for the whole range of the electric field considered.

\begin{figure}[t]
\includegraphics[scale=0.7]{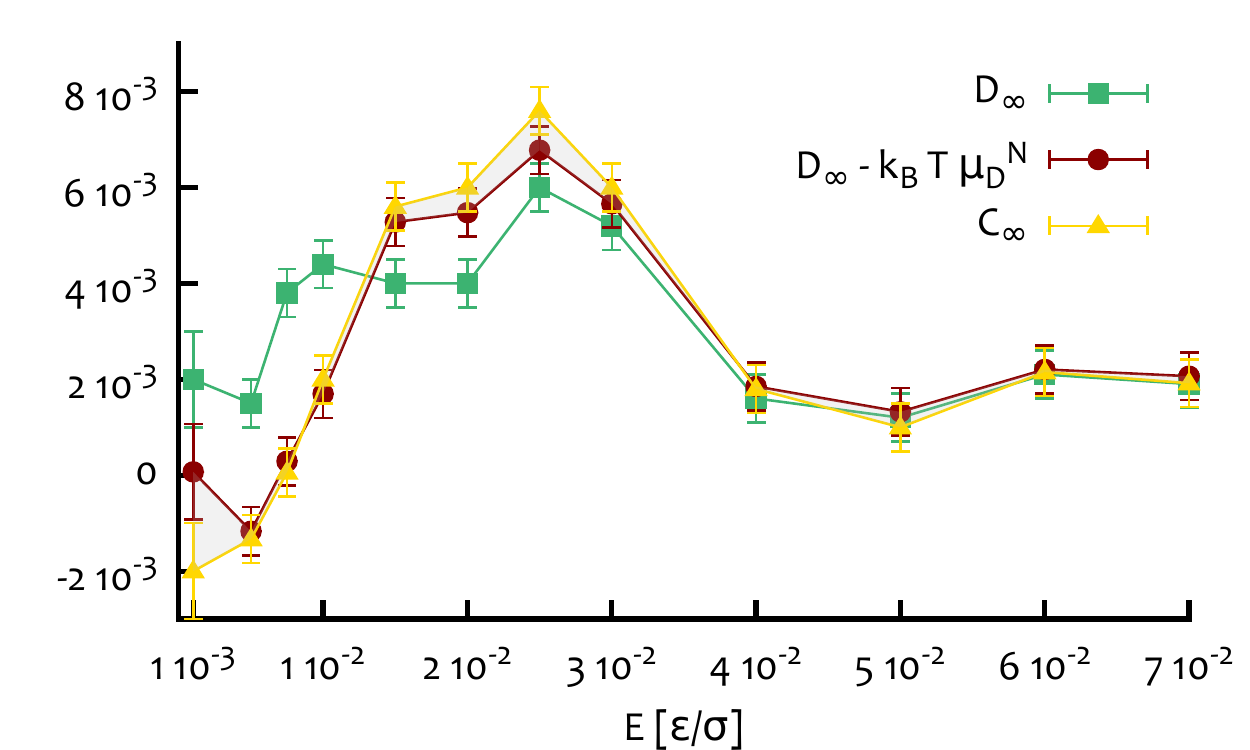}
\caption{The long time limit of the terms $\mathcal{D}$ (green), $\mathcal{D} - k_B T \mu_D^N$ (red) and $\mathcal{C}$ (yellow) are plotted against the force acting on single beads $qE$. While $\mathcal{D}$ and $\mu_D^N$ can be easily measured during an experiment~\cite{Mickel1977}, the latter ($\mathcal{C}$) can be retrieved by making use of eq.~\eqref{eq:DiffMobA} and Fig.~\ref{eq:DiffMobA}, \emph{i.e.} $\mathcal{C} = \mathcal{D} - k_BT \mu_D^N$. The shaded area highlights the small difference between the predicted value ($\mathcal{D} - k_B T \mu_D^N$) and the measured value of $\mathcal{C}$.}
\label{fig:DandC}
\end{figure}	

It is important to notice that  while tracking the position of particles travelling through a medium is a common experimental procedure, to get information on the forces acting on the particles is experimentally very difficult if not impossible. On the other hand, according to Eq.~\eqref{eq:DiffMobA} and Fig.~\ref{eq:DiffMobA}, if one knew $\mu_D^N$ and $\mathcal{D}$ from the experiments it would be possible to 
estimate $\mathcal{C}$ as~\cite{Bohec2013} \begin{equation}
\label{C_from_mu}
\mathcal{C} =  \mathcal{D} - k_B T \mu_D^N.
\end{equation}
Since $\mathcal{C}$ is the covariance between a total displacement and an average force (in the $z$ direction),
its value gives some information about the force experienced by a bead.

To show that the indirect determination of $\mathcal{C}$
in experiments is a viable strategy,
in Fig.~\ref{fig:DandC} we compare the long time limit values of $\mathcal{D} - k_B T \mu_D^N$ with values of $\mathcal{C}$ known directly from simulations, finding a very good agreement. 
 There is only a slight discrepancy between the $\mathcal{C}$ and $\mathcal{D} - k_B T \mu_D^N$ at very low fields, which is most likely due to errors in the measurements that become noisier at low fields, where diffusion dominates.
 
Note that by construction the generalized diffusion constant $\mathcal{D}>0$, also out of equilibrium~\cite{Baiesi2011}, see Fig.~\ref{fig:DandC}.
Conversely, one may also find negative covariances $\mathcal{C}$. Fig.~\ref{fig:DandC} suggests that this might happen for weak fields $E\gtrsim 0$. By increasing the field strength, however, the non-equilibrium term $\mathcal{C}$ rapidly increases until it overtakes the diffusive one, generating the regime of negative differential mobility. In such regime $\mathcal{D}$ can be much higher than in equilibrium and yet diffusion becomes secondary, as the nontrivial trapping of rings -- quantified by  $\mathcal{C}$ -- becomes the dominant phenomenon.

\section{Conclusions}
\label{s:conclusions}
In this paper we studied the role played by topology in the motion of ring polymers through a gel modelled as a disordered environment. We have shown that in this case, we can picture the gel structure as a collection of potential barriers, but only if polymers containing looped structures are moving though it. We simulated a system of linear polymers as a control system, and found that indeed these barriers do not represent a constraint on their migration. Our results therefore provide an explanation for irregular migration speeds detected in experiments comparing linearised and circular plasmids~\cite{Mickel1977,Levene1987}. In fact, our model predicts that ring polymers would show a negative differential mobility, \emph{i.e.} migrate slower and slower, for fields stronger than a certain value, in full agreement with experiments~\cite{Mickel1977}. The trapping of ring polymers by the environment is shown to be an effect as strong as the force drifting them. This is because the rings have to disengage from the dangling ends in order to re-establish their migration. This is equivalent to moving against an external bias and overcome a potential barrier whose height is proportional to the external bias itself. 

A different type of trapping is described by a self-threaded configuration~\cite{Michieletto2014c}, which has been recently shown to be a candidate for describing the low mobility of large circular polymers observed in pulse-field gel electrophoresis~\cite{Viovy1992}. In fact, a typical self-threaded configuration is much harder to disentangle even when pulse-field procedures are performed as it does not come undone if the direction of the external field is reversed. This is not the case for an impalement event. This means that in principle, one can distinguish impalements from self-threadings by looking at the mobility of DNA samples from gel electrophoresis experiments either at low fields or using a pulsed field~\cite{Levene1987}. In addition, our model can shed some light onto the problem of the irregular migration of knotted polymers in gel~\cite{Trigueros2001}, where the competition between the properties of the polymers due to their topology and the topological interaction with the gel plays an important role.

The phrase ``getting more from pushing less'' has recently been used to describe situations in which a higher current can be obtained by lowering the field strength~\cite{Zia2002,Baerts2013}. This phenomenon has been shown to well describe the behaviour of ellipsoidal Janus particles in corrugated channels~\cite{Ghosh2014}. Here, we presented another important realisation of this phenomenon, which is intimately related to the broadly used gel electrophoresis technique.  This is an instance of ``negative response'' and can be interpreted with recent advances in the field of non-equilibrium statistical mechanics. The covariance between displacement and average experienced force, which is null in equilibrium conditions, may become the dominant contribution to non-equilibrium susceptibilities (as the mobility) and overtake the diffusion term to render the response negative. 

In conclusion, with new theoretical results and above all with the proper choice of the polymeric probe, we conjecture that there is the possibility to explore the properties of complex molecular environments from fresh perspectives. In fact polymers with specific topologies can be exploited to design novel ways of sensing the changes in the microscopic structure of complex environments in a new and non-invasive way, \emph{e.g.} by looking at their mobilities.

\subsubsection*{Acknowledgment \hspace*{0.1 cm}  } 
DMi acknowledges the support from the Complexity Science Doctoral Training Centre at the University of Warwick with funding provided by the EPSRC (EP/E501311). EO acknowledge financial support from the Italian ministry of education grant PRIN 2010HXAW77. We also acknowledge the support of EPSRC to MST, EP/1005439/1, funding a Leadership Fellowship. The computing facilities were provided by the Centre for Scientific Computing of the University of Warwick with support from the Science Research Investment Fund.

\appendix 
\section*{Appendix}
\label{sec:appA}

Here, the formula for the non-equilibrium mobility $\mu_D^A$ is derived for one degree of freedom $z(t)$.
For simplicity, we consider over-damped dynamics, an approximation that applies well to the case we studied. The case of underdamped Langevin dynamics was also discussed in a previous work~\cite{Baiesi2011}. The over-damped equation of motion of a bead driven by the force $\Psi(z(t))$ and which is perturbed by the addition of a potential $-V(z) h(t)$ (where $h$ is a small parameter) and with inverse friction coefficient $\nu = 1/\xi$ at temperature $T$ ($k_B=1$) is 
\begin{equation}
\dot{z}(t) = \nu \left[ \Psi(z(t)) + h(t) \dfrac{\partial V}{\partial z}\right] + \sqrt{2 \nu T} \tilde{\eta}(t)
\end{equation}
where $\tilde{\eta}(t)$ is a standard white noise. Under quite general conditions, the response function for the observable $O(t)$ in overdamped systems~\cite{Baiesi2009c,Baiesi2011} is 
\begin{align}
R_{OV}(t,s) &= \left. \dfrac{\delta \langle O(t) \rangle^h}{\delta h(s)}\right|_{h=0} =  \\ 
&= \dfrac{1}{2T}\left[ \dfrac{d}{ds} \langle V(s)O(t) \rangle - \langle L V(s) O(t)\rangle \right],\notag
\end{align}
where $L$ is the backward generator of the Markov dynamics, 
\begin{equation}
L = \nu \Psi(z)\dfrac{\partial}{\partial z} + \nu T \dfrac{\partial^2}{\partial z^2}
\end{equation}
in this case.
The susceptibility 
\begin{align} \label{eq:susc}
\chi_{OV}(t) &= \int_0^t  R_{OV}(t,s) ds =  \\
&= \dfrac{1}{2 T} \left[  \langle \left[ V(t)-V(0)\right] O(t) \rangle - \left\langle \int_0^t  LV(s) ds\, O(t)\right\rangle \right].\notag
\end{align}
represents the linear response to a constant perturbation turned on at time $t=0$. 
In our case, the mobility is the susceptibility of the average velocity to the addition of a constant force,
\emph{i.e.} the observable is $O(t) = \Delta z(t)/t$ and the perturbation is $V(z) = z$, leading to $LV(z)=\nu\Psi(z)$. Hence \eqref{eq:susc} simplifies to  
\begin{equation}
\mu_D^A(t) = \dfrac{1}{2 T} \left[ \left\langle \Delta z(t) \dfrac{\Delta z(t)}{t} \right\rangle - \left\langle \int_0^t  \nu \Psi(s) ds \dfrac{\Delta z(t)}{t}\right\rangle \right].
\label{eq:mu1}
\end{equation}
Since we are out of equilibrium, the average displacement  $\langle \Delta z(t) \rangle$ is not zero. It is appropriate to remove it from the terms in eq.~\eqref{eq:mu1} and using $\langle \Delta z(t) \rangle = \left\langle \int_0^t \nu \Psi(s) ds \right\rangle$ we get
\begin{equation}
\mu_D^A(t) = \dfrac{1}{T} \left[ \dfrac{ \langle \Delta z(t);\, \Delta z(t) \rangle}{2 t} - \dfrac{1}{2 t}\left\langle \int_0^t \nu \Psi(s) ds ;\, \Delta z(t) \right\rangle \right]
\label{eq:mu2}
\end{equation}
where $\langle a; b\rangle \equiv \langle a b \rangle - \langle a \rangle \langle b \rangle$ denotes the covariance. The first term estimates the spread around the average position and can thus be interpreted as a diffusion constant, while the second represents a novel nonequilibrium term.  
The formula remains unaltered if the force $\Psi$ depends on many degrees of freedom, as long as their noises are statistically independent. The expression in \eqref{eq:mu2} is the one we used to compute the differential nonequilibrium mobility $\mu_D^A$ in eq.~\eqref{eq:DiffMobA}, letting $t$ become sufficiently large. Since beads are equivalent, we have also averaged the mobility of all beads.


\footnotesize{
\bibliography{EPRG_bib_OK} 
\bibliographystyle{rsc} 
}

\end{document}